 \documentclass[bm,aps,showpacs,amsmath,preprint,amssymb,12pt]{revtex4}

\usepackage{graphicx}
\usepackage{bm}

\begin{document}
{~}
\title{
Uniqueness theorem for charged rotating black holes in 
five-dimensional minimal supergravity
}

\vspace{2cm}

\author{Shinya Tomizawa}

\vspace{2cm}
\affiliation{
Cosmophysics Group, Institute of Particle and Nuclear Studies, 
KEK, 1-1, Oho, Tsukuba,  Ibaraki, 305-0801, Japan
}

\begin{abstract} 
We show a uniqueness theorem for charged rotating black holes in 
the bosonic sector of five-dimensional minimal supergravity. 
More precisely, under the assumptions of the existence of two commuting axial 
isometries and spherical topology of horizon cross-sections, we prove that 
an asymptotically flat, stationary charged rotating black hole with finite 
temperature in five-dimensional Einstein-Maxwell-Chern-Simons theory 
is uniquely characterized by the mass, charge, and two independent angular 
momenta and therefore is described by 
the five-dimensional Cveti{\v c}-Youm solution with equal charges.  
We also discuss a generalization of our uniqueness theorem for 
spherical black holes to the case of black rings. 

\end{abstract}

\date{\today}
\maketitle

\section{Introduction}\label{sec:intro}
In string theory and various related contexts, 
higher dimensional black holes and other extended black objects have played  
an important role. In particular, physics of black holes in five-dimensional 
Einstein-Maxwell-Chern-Simons (EMCS) theory has recently been the subject 
of increased attention, as the five-dimensional EMCS theory describes 
the bosonic sector of five-dimensional minimal supergravity, 
a low-energy limit of string theory. Various types of black hole solutions 
in EMCS-theory~\cite{Gauntlett0,BMPV,CY96,CLP,CCLP,Bena2,EEF,Elvang,EEMR,Elvang3,Gaiotto,IM,IKMT,IKMT2,NIMT,TIMN,MINT,TI,TYM,GS,TIKM,T} have so far been found, 
with the help of, in part, recent development of solution generating 
techniques~\cite{Belinskii,solitonbook,Tomizawa,Tomizawa2,Tomizawa3,Tomizawa4,Iguchiboost,Pomeransky:2005sj,Koikawa,Azuma,MishimaIguchi,Iguchi,EK07,CAM,BCCGSW,GS,Yaza06a,Yaza06b,Yaza08}. 
However, the classification of those black hole solutions has not been 
achieved yet. 
%
The purpose of this paper is to show a uniqueness theorem for charged 
rotating black holes in five-dimensional EMCS theory, as a partial solution 
to the black hole classification problem in string theory.

\medskip 
It is now evident that even within the framework of 
vacuum Einstein gravity, there is a much 
richer variety of black hole solutions in higher dimensions~\cite{
Myers:1986un,Emparan:2001wn,MishimaIguchi,Pom,diring,saturn,Izumi,bi}, 
the classification of which still remains a major open issue. 
As shown by Emparan and Reall~\cite{Emparan:2001wn}, 
five-dimensional vacuum Einstein gravity admits the co-existence of 
a rotating spherical hole and two rotating 
rings with the same conserved charges, 
illustrating explicitly the non-uniqueness property in higher dimensions. 
However, it is possible to show type of uniqueness theorems 
for some restricted cases in which certain additional conditions are 
imposed on some parameters/properties, other than the global conserved 
charges. 
For example, restricting attention to static solutions, 
Gibbons {\it et al.}~\cite{shiromizu} showed that the only asymptotically 
flat, static vacuum black hole is the Schwarzschild-Tangherlini 
solution~\cite{T-schwarzschild}. For rotating case, 
by assuming the existence of two axial Killing symmetries and spherical 
topology of the event horizon, Morisawa and Ida~\cite{MI} succeeded in 
proving that five-dimensional asymptotically flat, stationary vacuum 
rotating black holes must be in the Myers-Perry family.  
Their theorem was recently generalized to a class of asymptotically flat 
solutions with non-spherical horizon topology~\cite{Hollands,MTY}.  
For other cases (such as cases including Maxwell-field), 
see~\cite{HY,R,R2,R3,R4,R5,R6,R7,HY08}.

\medskip
In this paper, we generalize the boundary value analysis of Morisawa and 
Ida~\cite{MI} performed in vacuum Einstein gravity to the case of the bosonic 
sector of five-dimensional minimal supergravity.  
We are concerned with stationary black hole spacetimes that are asymptotically 
flat in the standard sense: Namely, we demand that the exterior region 
of the black hole is globally hyperbolic, having a spherical spatial 
infinity, and that the metric and other physical fields, such as Maxwell 
field, fall-off in a certain manner at large distances. 
(The asymptotic fall-off conditions are given later.) 
Furthermore, for simplicity, we focus on the single black hole case, 
that is, the event horizon is connected. 
Then, we note that in five-dimensional EMCS theory 
the Cveti{\v c}-Youm black hole solution with equal charges~\cite{CY96} 
appears to be the most general 
such solutions that describe an asymptotically flat, stationary charged 
rotating black hole with spherical horizon topology, characterized by 
four conserved charges, i.e., the mass, two independent angular momenta, 
and electric charge, and that encompass the known asymptotically flat, 
spherical black hole solutions in a subclass of EMCS theory, such as 
the Myers-Perry solution~\cite{Myers:1986un}, in a certain limit. 
Thus, we wish to show the following theorem. 

\medskip 
\noindent
{\bf Theorem.} 
{\em 
Consider, in five-dimensional Einstein-Maxwell-Chern-Simons theory 
[given by eq.~(\ref{action}) below], a stationary charged rotating 
black hole with finite temperature that is regular on and outside 
the event horizon and asymptotically flat in the standard sense 
with spherical spatial infinity. 
If (1) the black hole spacetime admits, besides the stationary Killing 
vector field, two mutually commuting axial Killing vector fields so that 
the isometry group is ${\Bbb R}\times U(1)\times U(1)$ and 
(2) the topology of the horizon cross-sections is spherical, ${\rm S^3}$, 
and the topology of the black hole exterior region is 
${\Bbb R} \times \{ {\Bbb R}^4 \setminus {\Bbb B}^4 \}$, 
then the black hole spacetime is uniquely characterized by its mass, 
electric charge, and two independent angular momenta, and hence must be 
isometric to the Cveti{\v c}-Youm solution with equal charges.
} 

\medskip 
Before presenting our proof, we would like to make a few comments 
concerning the assumptions made in our theorem. 
In order to obtain global results, we need the symmetry-condition (1), 
which, in particular 
makes it possible to reduce five-dimensional minimal supergravity to 
a non-linear sigma model with certain symmetries as shown in~\cite{MO,MO2}. 
Since all known exact black hole solutions in higher dimensions 
admit multiple axial isometries, our additional symmetry-condition~(1) 
does not appear to be too restrictive. However, we should note that 
the rigidity theorem~\cite{HIW} (see also \cite{MI08}) in higher 
dimensions---which is recently shown to be applicable also to EMCS 
theory~\cite{HI}---only guarantees the existence of a single rotational 
isometry (provided the spacetime metric and other fields are real, analytic), 
and therefore at present, the condition~(1) is not yet fully justified. 
In this respect, note also that the possibility for higher-dimensional 
black holes with fewer isometries than ${\Bbb R} \times U(1)\times U(1)$ 
has been suggested~\cite{Reall03}. 
Since the rigidity theorem yields that the event horizon is a Killing horizon, 
the notion of surface gravity is well-defined. Then, by 
{\em finite temperature} we mean that the event horizon 
is of non-degenerate type, having non-vanishing surface gravity 
and a bifurcate surface~\cite{RW92,RW96}. 
For extremal (zero-temperature) black holes with vanishing surface gravity,   
the event horizon is of degenerated type and does not possess a bifurcate 
surface. Then, our boundary conditions to be imposed on target space fields 
at the event horizon would not appear to straightforwardly apply to such a  
case that the horizon has no bifurcate surface. 
It would be of great interest to consider 
the classification problem of such extremal (zero-temperature) black holes. 
In this respect, there have recently appeared some attempts to classify 
near-horizon geometries of extremal black objects, rather than extremal 
black objects themselves (see e.g., \cite{KL08a,KL08b,FKLR08,KLR07,AGJST06,AY08} and references therein). 

\medskip 
We also need to additionally impose the topology-condition~(2), in order to 
explicitly specify boundary conditions on target space variables 
at the event horizon, in terms of certain coordinates, 
globally defined over the black hole exterior region. 
%
The topological censorship, together with our assumption of asymptotic 
flatness described above, immediately implies that the exterior region is 
topologically ${\Bbb R} \times V^{(4)}$ with $V^{(4)}$ being some 
four-dimensional simply connected Riemannian manifold. However, 
the simple connectedness by itself does not completely determine the topology 
of $V^{(4)}$. Therefore, in the present theorem, we simply demand that 
$V^{(4)} \approx \{{\Bbb R}^4 \setminus {\Bbb B}^4 \}$, which is in accordance 
with the topology of the Cveti{\v c}-Youm solution with equal charges. 
Our boundary conditions---in particular, the {\em rod structure}, 
which was first introduced by Harmark \cite{Harmark} based on earlier work 
for static solutions \cite{weyl}---are accordingly specified in the manner 
discussed in Sec.~\ref{sec:coincidence}.   
The topology theorem~\cite{Cai,Helfgott,galloway} yields that 
in five-dimensions, cross-sections of the event horizon must be 
topologically either a sphere, a ring, or a lens-space. 
The requirement (2) excludes some interesting class of solutions 
to be dealt with.  
It would be interesting to consider generalization of our uniqueness 
theorem to include solutions with non-spherical horizon topology. 

\medskip 
We would like to emphasize that even under these restrictive assumptions 
(1) and (2), still it is not at all obvious whether black holes in EMCS-theory 
are uniquely specified by their global charges. 
In fact, it has been shown by numerical studies~\cite{Kunz} that 
when the value of the Chern-Simons coupling is larger than some critical value,
spherical black holes in such a general EMCS-theory 
no longer enjoy the uniqueness property. 
In the present paper, motivated from sting theory, we restrict attention 
to a special class of EMCS-theory, that is, five-dimensional minimal 
supergravity and then are able to show the above uniqueness theorem.  
It would be interesting to find the precise onset of this non-uniqueness 
property in general EMCS theory, using the formulas developed in this paper.

\medskip
The rest of the paper is devoted to prove the above uniqueness theorem. 
In the next section, we present the metric and the gauge potential in 
Einstein-Maxwell-Chern-Simons theory with three Killing symmetries, 
introduce the Weyl-Papapetrou coordinates, and reduce the system 
to a non-linear sigma model with certain symmetries. 
In Sec.~\ref{sec:mazur}, using the matrix representation of the sigma model, 
we derive a divergence identity/Mazur identity associated with our nonlinear 
sigma model. 
A good part of the material in Sec.~\ref{sec:metric} and the first part 
of Sec.~\ref{sec:mazur} concerning the matrix representation 
is discussed in~\cite{BCCGSW}. 
Then, in Sec.~\ref{sec:coincidence}, presenting our boundary conditions 
for our sigma model fields and using the Mazur identity, we show that 
if two asymptotically flat black hole solutions have 
the same conserved charges, i.e., the mass, electric charge, and 
two angular momenta, then they must coincide with each other, and complete 
our proof of the uniqueness theorem. 
In Sec.~\ref{sec:conclusion}, we summarize our results and discuss possible 
generalization of our theorem to include non-spherical black objects. 
We discuss that in order to have a uniqueness theorem for black ring 
solutions in EMCS-theory, we need to specify rod-data, 
besides global charges and horizon topology. 
In Appendix~\ref{sec:A}, we explicitly compute relevant components of 
the Maxwell-field. 
In Appendix~\ref{sec:solution}, we provide the five-dimensional  
Cveti{\v c}-Youm solution with equal charges, and study, in terms of the Weyl-Papapetrou 
coordinates, the limiting behavior of the solution near relevant boundaries.

\section{Einstein-Maxwell-Chern-Simons system with symmetries}
\label{sec:metric}

We consider the bosonic sector of five-dimensional minimal supergravity 
theory, which can be obtained by a suitable truncation of eleven-dimensional 
supergravity. The five-dimensional action is given by 
\begin{eqnarray}
S=\frac{1}{16\pi}
  \left[ 
        \int dx^5\sqrt{-g}\left(R-\frac{1}{4}F^2\right) 
       -\frac{1}{3\sqrt{3}} \int F\wedge F\wedge A 
  \right] \,, 
\label{action} 
\end{eqnarray} 
where we set a Newton constant to be unity and $F=dA$. 
Varying this action (\ref{action}), we derive the Einstein equation 
\begin{eqnarray}
 R_{\mu \nu } -\frac{1}{2} R g_{\mu \nu } 
 = \frac{1}{2} \left( F_{\mu \lambda } F_\nu^{ ~ \lambda } 
  - \frac{1}{4} g_{\mu \nu } F_{\rho \sigma } F^{\rho \sigma } \right) \,, 
 \label{Eineq}
\end{eqnarray}
and the Maxwell equation 
\begin{eqnarray}
 d*F+\frac{1}{\sqrt{3}}F\wedge F=0 \,. 
\label{Maxeq}
\end{eqnarray}
The purpose of this section is to reduce the above five-dimensional 
Einstein-Maxwell-Chern-Simons system to a non-linear sigma model with 
certain target space symmetries. We first consider consequences of 
the existence of commuting Killing vector fields in our spacetime and 
identify the target space variables in Subsec.~\ref{subsec:II-A}. 
Then, having another (stationary) Killing vector field, we introduce 
the Weyl-Papapetrou coordinates and write down explicitly the desired 
non-linear sigma model action in Subsec.~\ref{subsec:II-B}.

\subsection{Two Killing system}\label{subsec:II-A}

Let $\xi_a \ (a=1,2)$ be two mutually commuting Killing vector fields,  
so that $[\xi_a,\xi_b]=0$, ${\cal L}_{\xi_a} g=0$, and ${\cal L}_{\xi_a}F=0$. 
Then, introducing the coordinates $x^a$ as Killing parameters of $\xi_a$ 
(so that $\xi_a = \partial/\partial x^a$), one can express the metric $g$ 
and the gauge potential one-form $A$, respectively, as     
\begin{eqnarray}
ds^2 = \lambda_{ab}(dx^a+a^a{}_idx^i)(dx^b+a^b{}_jdx^j) 
      +|\tau|^{-1}h_{ij}dx^idx^j \,, 
\end{eqnarray}
\begin{eqnarray} 
 A = A_adx^a + A_i dx^i \,, 
\end{eqnarray}
where the functions $\tau:=-{\rm det}(\lambda_{ab})$, $a^a{_i}$, $h_{ij}$, 
$A_a$, and $A_i$ ($i=3,4,5$) are independent of 
the coordinates~$x^a$.  

\medskip 
Let us define the electric one-form $E_a$ with respect to $\xi_a$ by 
\begin{eqnarray}
&&E_a=-i_{\xi_a}F \,.
\end{eqnarray}
Then the exterior derivatives of the electric one-forms yield
\begin{eqnarray}
dE_a=i_{\xi_a}dF-{\cal L}_{\xi_a}F=0 \,,
\end{eqnarray}
where $F=dA$ is used. Hence there exist locally the potentials $\psi_a$ 
such that 
\begin{eqnarray}
d\psi_a=-\frac{1}{\sqrt{3}}i_{\xi_a}F \,. 
\end{eqnarray} 
Hence, the gauge potential can be written as
\begin{eqnarray} 
 A = \sqrt{3}\psi_adx^a + A_i dx^i \,,
\end{eqnarray}
where $\psi_a$ is also independent of the coordinates $x^a$.
Next, define the magnetic one-form $B$ by 
\begin{eqnarray}
&&B=*(\xi_1\wedge \xi_2\wedge F) \,. 
\end{eqnarray}
Noting that $B$ can be rewritten as 
$
B=*(\xi_1\wedge \xi_2\wedge F)=-i_{\xi_2}*(\xi_1\wedge F)=i_{\xi_2}i_{\xi_1}*F
$ 
and using the identity 
$ 
di_{\xi_2}i_{\xi_1}=i_{\xi_2}i_{\xi_1}d+i_{\xi_1}{\cal L}_{\xi_2}-i_{\xi_2}{\cal L}_{\xi_1}
$, 
we can write the exterior derivative of $B$ as 
\begin{eqnarray}
dB=i_{\xi_2}i_{\xi_1}d*F \,. 
\end{eqnarray}
Then, using the Maxwell equation~(\ref{Maxeq}), we find that 
\begin{eqnarray}
dB= -\frac{1}{\sqrt{3}}i_{\xi_2}i_{\xi_1}F\wedge F 
  =  \frac{2}{\sqrt{3}} E_1\wedge E_2 
  = 2\sqrt{3} d\psi_1\wedge d\psi_2=\sqrt{3} d(\psi_1d\psi_2-\psi_2d\psi_1) \,.
\end{eqnarray}
This immediately implies that there exists the magnetic potential $\mu$ 
such that 
\begin{eqnarray}
d\mu=\frac{1}{\sqrt{3}}B-\epsilon^{ab}\psi_ad\psi_b \,,
\label{eq:mu} 
\end{eqnarray}
where $\epsilon^{12}=-\epsilon^{21}=1$.
We also introduce the twist one-form by 
\begin{equation} 
V_a=*(\xi_1\wedge \xi_2\wedge d\xi_a) \,. 
\end{equation}
Using the Einstein-equation, we can write the exterior derivative of $V_a$ as 
\begin{eqnarray}
dV_a &=& 2*(\xi_1\wedge \xi_2\wedge R(\xi_a)) \nonumber \\
    &=&-\tau^{-1}i_{\xi_2}i_{\xi_1}*^2(\xi_1\wedge\xi_2\wedge E_a\wedge B) 
\nonumber \\ 
    &=&-E_a\wedge B \nonumber \\
    &=&-3d\psi_a\wedge (d\mu+\epsilon^{bc}\psi_bd\psi_c) \nonumber \\
    &=&-3d[\psi_ad\mu]-d[\psi_a\epsilon^{bc}\psi_bd\psi_c] \,, 
\end{eqnarray}
where $R(\xi_a)$ in the first line is the Ricci one-form. 
Therefore, there exists the twist potentials $\omega_a$ that satisfy 
\begin{eqnarray}
 d\omega_a=V_a+\psi_a(3d\mu+\epsilon^{bc}\psi_bd\psi_c)\,.
\label{eq:twistpotential} 
\end{eqnarray}

\medskip
Thus, as a consequence of the existence of isometries $\xi_a$, 
we have eight scalar fields $\lambda_{ab},\omega_a,\psi_a,\mu$ 
$(a=1,2)$, which we denote collectively by coordinates 
$\Phi^A=(\lambda_{ab},\omega_a,\psi_a,\mu)$. As we will see soon, 
other components, such as $a^a{}_i$, $A_i$ are determined by $\Phi^A$.  
Then, we can find that the equations of motion, eqs.~(\ref{Eineq}) 
and (\ref{Maxeq}), are cast into a set of equations 
derived from the following action for sigma-model $\Phi^A$ 
coupled with three-dimensional gravity with respect to the metric $h_{ij}$, 
\begin{eqnarray}
S=\int_\Sigma\left({\cal R}^h
 -G_{AB}\frac{\partial \Phi^A}{\partial x^i}
 \frac{\partial \Phi^B}{\partial x^j}h^{ij}\right)\sqrt{|h|}dx^3 \,,
\end{eqnarray}
where the target space metric, $G_{AB}$, is given by 
\begin{eqnarray}
G_{AB}d\Phi^Ad\Phi^B 
&=& \frac{1}{4}{\rm Tr}(\lambda^{-1}d\lambda\lambda^{-1}d\lambda )
   + \frac{1}{4}\tau^{-2}d\tau^2 
   + \frac{3}{2}d\psi^T \lambda^{-1}d\psi \nonumber \\
& &-\frac{1}{2}\tau^{-1}V^T\lambda^{-1}V 
   -\frac{3}{2}\tau^{-1}(d\mu+\epsilon^{ab}\psi_ad\psi_b)^2 \,,
\end{eqnarray}
where $\lambda=(\lambda_{ab})$, $\psi=(\psi_1,\psi_2)^T$, $\omega=(\omega_1,\omega_2)^T$ and $V=d\omega-\psi(3d\mu+\epsilon^{bc}\psi_bd\psi_c)$.
Varying the action by $h_{ij}$, we obtain the equations
\begin{eqnarray}
R^h_{ij} = G_{AB}\frac{\partial \Phi^A}{\partial x^i} 
                 \frac{\partial \Phi^B}{\partial x^j} \,,  
\label{eq:Rij}
\end{eqnarray} 
where $R^h_{ij}$ denotes the Ricci tensor with respect to $h_{ij}$. 
Next varying the action by $\Phi^A$, we derive the equation
\begin{eqnarray}
\Delta_h\Phi^A+h^{ij}\Gamma^A_{BC}\frac{\partial \Phi^B}{\partial x^i} 
                                  \frac{\partial \Phi^C}{\partial x^j} 
  =0,
\end{eqnarray}
where $\Delta_h$ is the Laplacian with respect to the three-dimensional metric $h_{ij}$ and $\Gamma^A_{BC}$ is the Christoffel symbol with respect to the target space metric $G_{AB}$.

\subsection{Weyl-Papapetrou form}\label{subsec:II-B} 
Now we consider another Killing vector field $\xi_3$ 
which is assumed to commute with the other Killing vectors $\xi_a$ and will be 
identified below as the asymptotic time-translation Killing vector field. 
%
Let us consider the condition that the two-dimensional distribution orthogonal 
to three Killing vector fields $\xi_I\ (I=1,2,3)$ becomes integrable. 
The commutativity of Killing vector fields, $[\xi_I, \xi_J]=0$, enables 
us to find coordinate system $x_I$ $(I=1, 2, 3)$, so that 
$\xi_I=\partial/\partial {x^I}$ and the coordinate components of the metric 
become independent of $x^I$. 
We now recall the following theorem about the integrability of 
two-planes orthogonal to Killing vector fields~\cite{weyl,Harmark}: 

\medskip 
\noindent 
{\bf Proposition.}
{\em If three mutually commuting Killing vector fields $\xi_I\ (I=1,2,3)$ 
in a five-dimensional spacetime satisfy the following two conditions 
\begin{enumerate}
\item
$\xi_{1}^{[\mu _1}\xi_{2}^{\mu _2}
\xi_{3}^{\mu _{2}}D^\nu \xi_{I}^{\rho ]}=0$
holds at at least one point of the spacetime
for a given $I=1,2,3$, 

\item
$\xi_I^\nu R_\nu ^{[\rho}
\xi_{1}^{\mu _1}\xi_{2}^{\mu _2}
\xi_{3}^{\mu _{2}]}=0 $ holds for all
$I=1,2,3$,  
\end{enumerate} 
then the two-planes orthogonal to the Killing vector fields 
$\xi_I\ (I=1,2,3)$ are integrable.}

\medskip
Note here that one can replace a pair of Killing vector fields $(\xi_1,\xi_2)$ 
above by another pair $(\xi_2,\xi_3)$. We denote the corresponding 
quantities in the choice $(\xi_2,\xi_3)$ with {\em tilde} $\ \tilde{}\ $. 
For example, we denote the twist one-forms with respect to $(\xi_2,\xi_3)$ by 
\begin{eqnarray}
  \tilde V_{\tilde a}=*(\xi_2\wedge\xi_3\wedge d\xi_{\tilde a}) \,,
\end{eqnarray} 
where $\tilde a=2,3$.
Then, using 
$
i_{\xi_I}d\psi_a=i_{\xi_I}d\mu=0, \ i_{\xi_I}d\tilde\psi_a=i_{\xi_I}d\tilde\mu=0
$, 
and eq.~(\ref{eq:twistpotential}), we show 
$ 
i_{\xi_I}dV_a=i_{\xi_I}d\tilde V_{\tilde a}=0
$,   
and hence have 
\begin{eqnarray}
*(\xi_1\wedge \xi_2\wedge \xi_3\wedge R(\xi_a))
&=& -i_{\xi_3}*(\xi_1\wedge\xi_2\wedge R(\xi_a)) 
  = -\frac{1}{2}i_{\xi_3}dV_a =0 \,, 
\end{eqnarray}
and
\begin{eqnarray}
*(\xi_1\wedge \xi_2\wedge \xi_3\wedge R(\xi_3)) 
 &=& -i_{\xi_1}*(\xi_2\wedge\xi_3\wedge R(\xi_{3}))
  = -\frac{1}{2}i_{\xi_1}d\tilde V_{3}
 =0 \,. 
\end{eqnarray}
This implies that the condition~$2$ holds in our present 
system~(\ref{eq:action}) with three commuting Killing vector fields. 
Furthermore, the axial symmetry of at least one of $\xi_I$ $(I=1,2,3)$ 
implies that the condition~$1$ also holds on the axis of rotation. 
Therefore, the two-dimensional surface orthogonal to three $\xi_I$ is 
integrable.

\medskip 
Now, without loss of generality, we choose our three coordinates 
$(x^1,x^2,x^3)$ as the three Killing parameters, 
so that $\xi_3 = \partial/\partial t$ denotes the stationary 
(asymptotic time-translation) Killing vector field in our spacetimes and 
$\xi_1= \partial/ \partial \phi$ and $\xi_2 = \partial/\partial \psi$ 
are two independent axial Killing symmetries.  
Then, from the above observation, we can express the three dimensional metric 
$h_{ij}$ by $h=h_{pq}dx^pdx^q- \rho^2dt^2\ (p,q=4,5)$, where 
$\rho^2=-{\rm det}(g_{IJ})$. 
Note that the function $\rho$ is globally well-defined \cite{Chru08}. 
That $\rho$ is a harmonic function can be 
seen by looking at the $(tt)$-component of eq.~(\ref{eq:Rij}), 
which is written 
\begin{eqnarray}
 R_{tt}=\rho \hat D^2\rho=0 \,,
\end{eqnarray}
where $\hat D_p$ is the covariant derivative associated with 
the two-dimensional metric $h_{pq}$. 
Let $z$ be harmonic function conjugate to $\rho$ which satisfies 
$\hat D^2 z=0$, $\hat D_p \rho\hat D^pz=0$, 
$\hat D_p \rho\hat D^p\rho=\hat D_p z\hat D^p z$. 
Choose the coordinates $(x^4,x^5)$ as $x^4=\rho$ and $x^5=z$. 
Then, the metric can be written in the Weyl-Papapetrou type form as 
\begin{eqnarray}
ds^2&=&\lambda_{\phi\phi}(d\phi+a^\phi{}_tdt)^2+\lambda_{\psi\psi}(d\psi+a^\psi{}_tdt)^2 \nonumber\\
&&+2\lambda_{\phi\psi}(d\phi+a^\phi{}_tdt)(d\psi+a^\psi{}_tdt)+|\tau|^{-1}[e^{2\sigma}(d\rho^2+dz^2)-\rho^2 dt^2] \,, 
\end{eqnarray}
where all the metric components depend only on $\rho$ and $z$. 

\medskip 
In this coordinate system, $\Phi^A$ are determined by 
the equations of motion 
\begin{eqnarray}
\Delta_\gamma \Phi^A + 
\Gamma^A_{BC}[\Phi^B_{,\rho}\Phi^C_{,\rho}+\Phi^C_{,z}\Phi^C_{,z}]=0 \,, 
\label{eq:scalar}
\end{eqnarray}
where $\Delta_\gamma$ is the Laplacian with respect to the abstract 
three-dimensional metric $\gamma=d\rho^2+dz^2+\rho^2d\varphi^2$. 
On the other hand, once $\Phi^A$ are given, one can completely determine 
$\sigma$, $a^\phi{}_t$, $a^\psi{}_t$, $A_i$. 
In fact, the function $\sigma$ is determined by 
\begin{eqnarray}
\frac{2}{\rho}\sigma_{,\rho}&=&R^h_{\rho\rho}-R^h_{zz}\nonumber\\
 &=&G_{AB}[\Phi^A_{,\rho}\Phi^B_{,\rho}-\Phi^A_{,z}\Phi^B_{,z}] \,,\\
   \frac{1}{\rho}\sigma_{,z}&=&R^h_{\rho z}\nonumber\\
                            &=&G_{AB}\Phi^A_{,\rho}\Phi^B_{,z} \,.
\end{eqnarray}
The integrability $\sigma_{,\rho z}=\sigma_{,z \rho }$ is assured by  
eq.~(\ref{eq:scalar}). From eq.(\ref{eq:twistpotential}), the metric functions $a^a{_t}$ are determined by 
\begin{eqnarray}
a^a{}_{t,\rho}
&=&\rho \tau^{-1} \lambda^{ab}(\omega_{b,z}-3\psi_b\mu_{,z}-\psi_b\epsilon^{cd}\psi_c\psi_{d,z}) \\
a^a{}_{t,z}&=&-\rho\tau^{-1}\lambda^{ab}(\omega_{b,\rho}-3\psi_b\mu_{,\rho}-\psi_b\epsilon^{cd}\psi_c\psi_{d,\rho}) .
\end{eqnarray}
As shown in Appendix~\ref{sec:A}, we can set $A_{\rho}=A_{z}=0$.
Therefore it follows from eq.~(\ref{eq:mu}) that 
the $t$-component of the gauge potential $A$ is determined by 
\begin{eqnarray}
A_{t,\rho}&=&\sqrt{3}\left[a^a{}_t\psi_{a,\rho}-\rho \tau^{-1}(\mu_{,z}+\epsilon^{bc}\psi_b\psi_{c,z})\right],\\
A_{t,z}&=&\sqrt{3}\left[a^a{}_t\psi_{a,z}+\rho\tau^{-1}(\mu_{,\rho}+\epsilon^{bc}\psi_b\psi_{c,\rho})\right].
\end{eqnarray}
Thus, once we determine $\Phi^A= (\lambda_{ab},\omega_a,\psi_a,\mu)$, 
we can specify the solutions of the system given originally 
by the action, eq.~(\ref{action}), with our Killing symmetry assumption.  
It turns out that the above equations of motion, eq.~(\ref{eq:scalar}), 
for $\Phi^A$ are derived from the following action
\begin{eqnarray}
S&=&\int d\rho dz \rho\left[ G_{AB}(\partial\Phi^A)(\partial\Phi^B)\right] 
\nonumber \\
 &=& \int d\rho dz \rho 
     \biggl[\:  
             \frac{1}{4}{\rm Tr}(\lambda^{-1}\partial\lambda\lambda^{-1}
                                             \partial\lambda )
           + \frac{1}{4}\tau^{-2}\partial\tau^2 
           + \frac{3}{2}\partial\psi^T \lambda^{-1}\partial\psi 
\nonumber \\
& &{} \qquad \qquad 
           - \frac{1}{2}\tau^{-1}v^T\lambda^{-1}v 
           - \frac{3}{2}\tau^{-1}(\partial\mu 
                                   +\epsilon^{ab}\psi_a\partial\psi_b)^2 
     \biggr]  \,,
\label{eq:action} 
\end{eqnarray} 
where $v=\partial \omega-\psi(3\partial \mu+\epsilon^{bc}\psi_b\partial \psi_c)$.
This action is invariant under the global $G_{2(2)}$ transformation.

\section{Mazur identity}\label{sec:mazur} 

In the proof of uniqueness theorems for four-dimensional charged 
rotating black holes, a key role was played by a certain global 
identity---called the Mazur identity. 
This is also the case for five-dimensional charged rotating black holes. 
In this section, we present the Mazur type identity for our non-linear 
sigma models derived in the previous section. 
The derivation parallels that for the vacuum Einstein case given in other 
literature, e.g., Morisawa and Ida~\cite{MI}, and therefore we present here 
only some key formulas.   

\medskip 
Following ~\cite{BCCGSW}, we introduce the $G_{2(2)}/SO(4)$ coset 
matrix, $M$, defined by 
\begin{eqnarray}
M= \left(
  \begin{array}{ccc}
  \hat A&\hat B&\sqrt{2}\hat U\\
  \hat B^T&\hat C&\sqrt{2}\hat V\\
  \sqrt{2}\hat U^T&\sqrt{2}\hat V^T&\hat S\\
  \end{array}
 \right) \,,
\end{eqnarray}
where $\hat A$ and $\hat C$ are symmetric $3\times 3$ matrices, $\hat B$ is a $3\times 3$ 
matrix, $\hat U$ and $\hat V$ are 3-component column matrices, and $\hat S$ is a scalar, 
defined, respectively, by  
\begin{eqnarray}
&&\hat A=\left(
  \begin{array}{ccc}
  [(1-y)\lambda+(2+x)\psi \psi^T-\tau^{-1}\tilde\omega\tilde\omega^T+\mu(\psi \psi^T\lambda^{-1}\hat J-\hat J\lambda^{-1}\psi\psi^T)]&\tau^{-1}\tilde\omega\\
 \tau^{-1}\tilde\omega^T& -\tau^{-1}
  \end{array}
 \right) \,,\nonumber\\
&&\hat B=\left(
  \begin{array}{ccc}
  (\psi\psi^T-\mu \hat J)\lambda^{-1}-\tau^{-1}\tilde\omega \psi^T \hat J&[(-(1+y)\lambda \hat J-(2+x)\mu+\psi^T\lambda^{-1}\tilde\omega)\psi+(z-\mu \hat J\lambda^{-1}\tilde)\omega] \\
  \tau^{-1}\psi^T \hat J&-z\\
  \end{array}
 \right) \,, \nonumber\\
&&\hat C=\left(
  \begin{array}{ccc}
 (1+x)\lambda^{-1}-\lambda^{-1}\psi\psi^T\lambda^{-1}&\lambda^{-1}\tilde\omega-\hat J(z-\mu \hat J \lambda^{-1})\psi\\
 \tilde\omega^T\lambda^{-1}+\psi^T(z+\mu \lambda^{-1}\hat J)\hat J&[\tilde\omega^T\lambda^{-1}\tilde\omega-2\mu\psi^T\lambda^{-1}\tilde\omega-\tau(1+x-2y-xy+z^2)]\\
  \end{array}
 \right) \,,\nonumber\\
&&\hat U=
\left(
  \begin{array}{cc}
  (1+x-\mu \hat J\lambda^{-1})\psi-\mu \tau^{-1}\tilde\omega\\
  \mu\tau^{-1}\\
  \end{array}
 \right) \,,\nonumber\\
&&\hat V=\left(
  \begin{array}{ccc}
  (\lambda^{-1}+\mu\tau^{-1}\hat J)\psi\\
  \psi^T\lambda^{-1}\tilde\omega-\mu(1+x-z)\nonumber\\
  \end{array}
 \right) \,,\nonumber\\
&&\hat S=1+2(x-y) \,, \nonumber
\end{eqnarray}
with 
\begin{eqnarray}
&&\tilde \omega=\omega-\mu\psi \,,
\end{eqnarray}
\begin{eqnarray}
&&x=\psi^T\lambda^{-1}\psi,\quad y 
   =\tau^{-1}\mu^2,\quad z=y-\tau^{-1}\psi^T\hat J\tilde\omega \,,
\end{eqnarray}
and the $2\times 2$ matrix, 
\begin{eqnarray}
\hat J= \left(
  \begin{array}{ccccccc}
   0&1\\
   -1&0\\
  \end{array}
 \right) \,.  
\end{eqnarray}
We note that this $7\times7$ matrix $M$ is symmetric, $M^T=M$, 
and unimodular, $\det(M)= 1$. Since we choose the Killing vector fields 
$\xi_\phi$ and $\xi_\psi$ to be spacelike, all the eigenvalues of $M$ are 
real and positive. Therefore, there exists an $G_{2(2)}$ matrix $\hat g$ such that 
\begin{eqnarray}
  M = \hat g \hat g^T \,.
\end{eqnarray} 
We define a current matrix as
\begin{eqnarray}
 J_i = M^{-1} \partial_i M \,,
\end{eqnarray}
which is conserved if the scalar fields are the solutions of the equation of motion derived by the action (\ref{eq:action}). Then, the action (\ref{eq:action}) can be written in terms of $J$ and $M$ as follows
\begin{eqnarray}
S&=&\frac{1}{4}\int d\rho dz \rho {\rm tr}(J_iJ^i) \nonumber \\
 &=&\frac{1}{4}\int d\rho dz 
    \rho {\rm tr}(M^{-1}\partial_iMM^{-1}\partial^iM)\,. 
\end{eqnarray}
Thus, the matrix $M$ completely specify the solutions to our system.   

\medskip 
Let us now consider two sets of field configurations, 
$M_{[0]}$ and $M_{[1]}$, that satisfy the equations of motion derived 
from the action, eq.~(\ref{eq:action}). 
We denote the difference between the value of the functional obtained from 
the field configuration $M_{[1]}$ and the value obtained from $M_{[0]}$ 
as a bull's eye $\stackrel{\odot}{}$, e.g., 
\begin{eqnarray}
\stackrel{\odot}J{}^i=J^i_{[1]}-J^i_{[0]} \,,
\end{eqnarray}
where the subscripts ${}_{[0]}$ and ${}_{[1]}$ denote, respectively, 
the quantities associated with the field configurations $M_{[0]}$ 
and $M_{[1]}$. The deviation matrix, $\Psi$, is then defined by
\begin{eqnarray}
\Psi=\stackrel{\odot}MM^{-1}_{[0]}=M_{[1]}M^{-1}_{[0]}-{\bf 1} \,,
\end{eqnarray}
where ${\bf 1}$ is the unit matrix. Taking the derivative of this, 
we have the relation between the derivative of the deviation matrix 
and $\stackrel{\odot}J{}^i$, 
\begin{eqnarray}
D^i\Psi=M_{[1]}\stackrel{\odot}J{}^iM_{[0]}^{-1} \,, 
\label{eq:deriv}
\end{eqnarray}
where $D_i$ is a covariant derivative associated with the abstract 
three-metric $\gamma$. 
Taking, further, the divergence of the above formula and also the trace of 
the matrix elements, we have the following divergence identity 
\begin{eqnarray}
D_i D^i {\rm tr} \Psi 
 = {\rm tr} \left( \stackrel{\odot}J{}^{Ti} M_{[1]}\stackrel{\odot}J{}^iM_{[0]}^{-1} \right) \,,   
\label{id:global-divergence}
\end{eqnarray}
where we have also used the conservation equation $D_i J{}^i =0$. 
Then, integrating this divergence identity over the region 
$\Sigma=\{(\rho,z)|\rho\ge 0,\ -\infty<z<\infty \}$, we obtain 
the Mazur identity, 
\begin{eqnarray}
\int_{\partial \Sigma}\rho \partial_p {\rm tr} \Psi dS^p 
= \int_{\Sigma}\rho \hat h_{pq}{\rm tr} 
 ({\cal M}^{Tp} \: {\cal M}^q)d\rho dz \,, 
\label{eq:id} 
\end{eqnarray}
where $\hat h_{pq}$ is the two-dimensional flat metric 
\begin{eqnarray}
\hat h=d\rho^2+dz^2 \,, 
\end{eqnarray}
and the matrix ${\cal M}$ is defined by 
\begin{eqnarray}
{\cal M}^p=\hat g_{[0]}^{-1}\: \stackrel{\odot}J{}^{Tp} \: \hat g_{[1]} \,. 
\end{eqnarray}

\medskip
Now we note that the right-hand side of the identity, (\ref{eq:id}), 
is non-negative. 
Therefore, if we impose the boundary conditions at $\partial\Sigma$,
under which the left-hand side of Eq.(\ref{eq:id}) vanishes, then 
we must have $\stackrel{\odot}J{}^i=0$. In that case, 
it follows from eq.~(\ref{eq:deriv}) that $\Psi$ must be a constant matrix 
over the region $\Sigma$. 
Therefore, in particular, if $\Psi$ is shown to be zero on some part of 
the boundary $\partial \Sigma$, it immediately follows that $\Psi$ must 
be identically zero over the base space $\Sigma$, implying that 
the two solutions $M_{[0]}$ and $M_{[1]}$ must coincide with each other. 
This is indeed the case under our boundary conditions 
discussed in the next section.   


\section{Boundary value problems} \label{sec:coincidence}
In this section, we derive necessary boundary conditions for determining 
the scalar fields $\Phi^A=(\lambda_{ab},\omega_a,\psi_a,\mu)$, 
requiring asymptotic flatness at infinity, regularity on the two rotation 
axes (i.e., the $\phi$-invariant plane and the $\psi$-invariant plane),  
and on the event horizon (of which cross-sections are 
assumed to be topologically spherical). 
Note that by {\em asymptotically flat}, we mean that the spacetime 
metric has the following fall off behavior at large distances, 
\begin{eqnarray}
ds^2 &\simeq& \left(-1+\frac{8M_{ADM}}{3\pi r^2}+{\cal O}(r^{-3})\right)dt^2
            - \left(
                    \frac{8J_\phi\sin^2\theta}{\pi r^2}+{\cal O}(r^{-3})
              \right)dtd\phi\nonumber\\ 
       & &- \left(
                    \frac{8J_\psi\cos^2\theta}{\pi r^2}+{\cal O}(r^{-3})
              \right)dtd\psi 
\nonumber\\
      &   & + \left( 1 + {\cal O}(r^{-1}) \right)
                  \left(dr^2+
                        r^2\left(d\theta^2 + \sin^2\theta d\phi^2  
                                  + \cos^2\theta d\psi^2 \right)
                  \right) \,, 
\label{condi:AF}
\end{eqnarray} 
having the spherical spatial infinity, $S^3_\infty$.   
Here the constants $M_{ADM}$ and $J_a$ are the asymptotic conserved mass and 
angular momenta. Since we are concerned with stationary, axisymmetric 
spacetimes with Killing symmetries $\xi_I$, the conserved charges 
$M_{ADM}$ and $J_a$ are defined, respectively, by 
\begin{eqnarray} 
 M_{ADM} &=& -\frac{3}{32\pi} 
        \int_{S^3_\infty}
                 dS^{\mu \nu} \nabla_\mu (\xi_3)_\nu \,,  
\label{def:Mass} 
\\
 J_a &=& \frac{1}{16\pi}\int_{S^3_\infty} 
               dS^{\mu \nu} \nabla_\mu (\xi_a)_\nu \,. 
\label{def:Ja} 
\end{eqnarray}  
We write below our boundary conditions for $\Phi^A$ in terms of 
the Weyl-Papapetrou coordinates. Therefore, in particular, relevant 
conditions at infinity---see below eqs.~(\ref{gtt}) -- (\ref{grhorho})---are 
derived from the above fall-off behavior, eq.~(\ref{condi:AF}), 
by the coordinate transformation   
\begin{eqnarray}  
 \rho = \frac{1}{2} r^2 \sin 2\theta \,, \quad 
    z = \frac{1}{2} r^2 \cos 2\theta \,. 
\end{eqnarray}  
Then, we can find that the boundary conditions given in this section are, 
in fact, the same as the limiting behavior of $\Phi^A$ for the exact 
solution of Cveti\v{c}-Youm~\cite{CY96} at 
the corresponding boundaries, which we discuss in Appendix~\ref{sec:solution}. 

\medskip 
In terms of the Weyl-Papapetrou coordinate system introduced in 
Sec.~\ref{subsec:II-B}  
and the rod-structure \cite{Harmark}, 
the boundary $\partial \Sigma$ of the base space 
$\Sigma=\{(\rho,z)|\ \rho>0,\ -\infty<z<\infty \}$ is described as 
a set of three rods and the infinity: Namely, 
\begin{enumerate}
\item[(i)]
the $\phi$-invariant plane: 
$\partial \Sigma_\phi=\{(\rho,z)|\rho=0,k^2<z<\infty \}$ with the rod vector 
$v=(0,1,0)$ \,, 
\item[(ii)] the horizon: 
$\partial \Sigma_{\cal H}=\{(\rho,z)|\ \rho=0,-k^2<z<k^2\}$ \,, 
\item[(iii)] 
the $\psi$-invariant plane: 
$\partial \Sigma_\psi=\{(\rho,z)|\rho=0,-\infty<z<-k^2\}$ 
with the rod vector $v=(0,0,1)$ \,, 
\item[(iv)] the infinity:  
$\partial \Sigma_\infty 
= \{(\rho,z)|\sqrt{\rho^2+z^2}\to 
\infty\ {\rm with}\ z/\sqrt{\rho^2+z^2}\ {\rm finite}  \}$ \,, 
\end{enumerate} 
where here and hereafter ${\cal H}$ denotes a spatial cross-section of 
the event horizon. 
%
Accordingly, the boundary integral in the left-hand side of the Mazur 
identity, eq.~(\ref{eq:id}), is decomposed into the integrals over the three 
rods (i)--(iii), and the integral at infinity (iv), as 
\begin{eqnarray}
\int_{\partial \Sigma}\rho\partial_p{\rm tr}\Psi dS^p 
&=& \int_{-\infty}^{-k^2}\rho\frac{\partial {\rm tr}\Psi }{\partial z}dz 
   +\int_{-k^2}^{k^2}\rho\frac{\partial {\rm tr}\Psi }{\partial z}dz 
\nonumber\\
 && +\int_{k^2}^{\infty}\rho\frac{\partial {\rm tr}\Psi }{\partial z}dz 
    +\int_{\partial\Sigma_\infty}\rho\partial_a{\rm tr}\Psi dS^a \,.   
\label{eq:integral} 
\end{eqnarray}

\medskip
In order to evaluate this boundary integral, let us first consider 
the integrals of the twist one-forms $d\omega_a$ along the $z$-axis. 
By definition, the partial derivatives with respect to $z$ of the twist 
potentials $\omega_a$ vanish on both rotation axes. 
This means that the twist potentials $\omega_a$ are constant over 
the $\phi$-invariant plane and the $\psi$-invariant plane. 
Therefore, the integral can be written as 
\begin{eqnarray}
\int_{-\infty}^{\infty}\omega_{a,z}dz
 &=& \int_{-k^2}^{k^2}\omega_{a,z}dz \\ 
 &=& \biggl[\omega_a\biggr]_{z=-k^2}^{z=k^2} \,. 
\label{eq:int} 
\end{eqnarray}
On the other hand, by Stokes's theorem, the integral of $d\omega_a$ 
on the horizon is evaluated as 
\begin{eqnarray}
\int_{\partial\Sigma_{\cal H}}d\omega_a 
   &=& \int_{\partial\Sigma_\infty}d\omega_a\nonumber \\ 
   &=& \int_{\partial\Sigma_\infty}V_a 
    +\int_{\partial\Sigma_\infty}\psi_a(3d\mu+\epsilon^{bc}\psi_bd\psi_c) \,. 
\label{eq:int3}
\end{eqnarray}
We find that the first integral in the right-hand side of eq.~(\ref{eq:int3}) 
is proportional to the angular momenta $J_a$, defined by eq.~(\ref{def:Ja}) 
above. 
As will be seen later, the second integral vanishes at infinity. 
Hence, using the degrees of freedom in adding a constant to $\omega_a$, 
we can always set the value of $\omega_a$ on the two rotation axes to be  
\begin{eqnarray}
\omega_a(z)=-\frac{2J_a}{\pi} \,, \label{eq:twist}
\end{eqnarray}
for $z\in [k^2,\infty]$, and
\begin{eqnarray}
\omega_a(z)=\frac{2J_a}{\pi} \,, \label{eq:twist2} 
\end{eqnarray}
for $z\in [-\infty,-k^2]$.

\medskip
Next, consider the integral of $\mu_{,z}$ on the horizon 
$\partial \Sigma_{\cal H}$. The derivative of the potential, $d\mu$, vanishes 
on the two rotation axes by definition. Hence the integral along the $z$-axis 
becomes 
\begin{eqnarray}
\int^{\infty}_{-\infty}\mu_{,z}dz=\int^{k^2}_{-k^2}\mu_{,z}dz=\biggl[\ \mu\ \biggr]_{z=-k^2}^{z=k^2}.
\end{eqnarray}
We find that this integral is proportional to the electric charge $Q$ 
defined by 
\begin{eqnarray}
Q&=&\frac{1}{16\pi}\int_{\cal H}\left(*F+\frac{1}{\sqrt{3}}A\wedge F\right) \,. 
\end{eqnarray} 
In fact, straightforward calculation shows 
\begin{eqnarray}
&&\frac{1}{16\pi}\int_{\cal H}\left(*F+\frac{1}{\sqrt{3}}A\wedge F\right)\nonumber\\ 
 &=&\frac{\pi}{4}\int^{k^2}_{-k^2}
    \left[
          \frac{\tau}{\rho}(A_{t,\rho}-a^\phi{}_tA_{\phi,\rho}
          -a^\psi{}_tA_{\psi,\rho}) 
          -\frac{1}{\sqrt{3}}(A_\phi A_{\psi,z}-A_\psi A_{\phi,z}) 
    \right]dz 
\nonumber \\
 &=&\frac{\pi\sqrt{3}}{4}\int^{k^2}_{-k^2} \mu_{,z}dz \,.  
\end{eqnarray}
Hence, without loss of generality, $\mu$ can be set to be 
\begin{eqnarray} 
\mu=-\frac{2Q}{\sqrt{3}\pi} \,, 
\label{eq:charge1}
\end{eqnarray}
for $\rho=0,\ z\in [-\infty,-k^2]$, and 
\begin{eqnarray}
\mu=\frac{2Q}{\sqrt{3}\pi} \,, 
\label{eq:charge2} 
\end{eqnarray}
for $\rho=0,\ z\in [k^2,\infty]$.

\medskip 
Now we would like to show that the boundary integral, 
eq.~(\ref{eq:integral}), indeed vanishes under our preferable boundary 
conditions that require the regularity on the three rods and asymptotic 
flatness at infinity. For this purpose, in the following we evaluate 
the limiting behavior of the integrand, $\rho\ \partial_z{\rm tr}\ \Psi$, 
of eq.~(\ref{eq:integral}), separately on each boundary (i)--(iv).  

\smallskip 
\noindent 
(i) $\phi$-invariant plane: 
$\partial\Sigma_\phi=\{(\rho,z)|\rho=0,\ k^2<z<\infty\}$. 
The regularity on the $\phi$-invariant plane requires that for $\rho\to 0$, 
the scalar fields behave as 
\begin{eqnarray}
&&\lambda_{\phi\phi}\simeq {\cal O}(\rho^2) \,,\label{eq:a1}\\ 
&&\lambda_{\psi\psi}\simeq {\cal O}(1) \,,\label{eq:a2}\\
&&\lambda_{\phi\psi}\simeq {\cal O}(\rho^2) \,, \label{eq:a3}\\
&&\omega_{\phi}\simeq -\frac{2J_\phi}{\pi}+{\cal O}(\rho^2) \,, \label{eq:a4}\\
&&\omega_{\psi}\simeq -\frac{2J_\psi}{\pi}+{\cal O}(\rho^2) \,,\label{eq:a5} 
\end{eqnarray}
and
\begin{eqnarray}
&&\psi_{\phi}\simeq {\cal O}(\rho^2)\,, \label{eq:a6}\\
&&\psi_{\psi}\simeq {\cal O}(1) \,,\label{eq:a7}\\
&&\mu\simeq \frac{2Q}{\sqrt{3}\pi}+{\cal O}(\rho^2) \,, \label{eq:a8} 
\end{eqnarray}
where the boundary conditions, eqs.~(\ref{eq:a1})-(\ref{eq:a3}) and 
eqs.~(\ref{eq:a6})-(\ref{eq:a7}), come from the requirement that 
$\partial\Sigma_\phi$ is the $\phi$-invariant plane, i.e., 
the plane invariant under the rotation with respect to the axial 
Killing vector $\partial/\partial \phi$. 
The conditions, eqs.~(\ref{eq:a4})-(\ref{eq:a5}), are derived from 
eq.~(\ref{eq:twist}). 
In the derivation of the condition (\ref{eq:a8}), eq.~(\ref{eq:charge2}) 
is used. 
Then for two solutions, $M_{[0]}$ and $M_{[1]}$, with the same mass, 
the same angular momenta, and the same electric charge, 
$\rho\ {\rm tr}\Psi$ behaves as 
\begin{eqnarray}
\rho\ \partial_z{\rm tr}\ \Psi\simeq O(\rho) \,. 
\end{eqnarray}

\smallskip 
\noindent 
(ii) Horizon: $\partial \Sigma_{\cal H}=\{(\rho,z)|\ \rho=0, \ -k^2<z<k^2 \}$.
The regularity on the horizon requires that 
for $\rho \to 0$,   
\begin{eqnarray}
&&\lambda_{ab}\simeq {\cal O}(1),\quad \omega_{a}\simeq {\cal O}(1) \,,\\
&&\psi_{a}\simeq {\cal O}(1), \quad \mu\simeq {\cal O}(1) \,. 
\end{eqnarray}
Therefore, for $\rho \to 0$, $\rho\ {\rm tr}\Psi$ behaves as
\begin{eqnarray}
\rho\ \partial_z{\rm tr}\ \Psi\simeq O(\rho) \,.  
\end{eqnarray}

\smallskip 
\noindent 
(iii) $\psi$-invariant plane: 
$\partial\Sigma_\psi=\{(\rho,z)|\rho=0,\ -\infty<z<-k^2\}$. 
Similarly to the case (i), the regularity on the $\phi$-invariant plane 
requires 
\begin{eqnarray}
&&\lambda_{\psi\psi}\simeq {\cal O}(\rho^2) \, ,\label{eq:b1}\\
&&\lambda_{\phi\phi}\simeq {\cal O}(1) \,, \label{eq:b2}\\
&&\lambda_{\phi\psi}\simeq {\cal O}(\rho^2) \,, \label{eq:b3}\\
&&\omega_{\phi}\simeq \frac{2J_\phi}{\pi}+{\cal O}(\rho^2) \,, \label{eq:b4}\\
&&\omega_{\psi}\simeq \frac{2J_\psi}{\pi}+{\cal O}(\rho^2) \,, \label{eq:b5}
\end{eqnarray}
and 
\begin{eqnarray}
&&\psi_{\phi}\simeq {\cal O}(1) \,, \label{eq:b6} \\
&&\psi_{\psi}\simeq {\cal O}(\rho^2) \,,\label{eq:b7} \\
&&\mu\simeq -\frac{2Q}{\sqrt{3}\pi}+{\cal O}(\rho^2) \,. \label{eq:b8}
\end{eqnarray}
Therefore, for $\rho \to 0$, $\rho\ {\rm tr}\Psi$ behaves as
\begin{eqnarray}
\rho\ \partial_z{\rm tr}\ \Psi \simeq O(\rho) \,. 
\end{eqnarray}

\smallskip 
\noindent 
(iv) Infinity: 
$\partial\Sigma_\infty=\{(\rho,z)|\ \sqrt{\rho^2+z^2}\to\infty$ with 
$z/\sqrt{\rho^2+z^2}$ finite $\}$. Recall that the three-dimensional metric 
$g=(g_{IJ})\ (I,J=t,\phi,\psi)$ is subject to the constraint 
\begin{eqnarray}
{\rm det}(g)=-\rho^2 \,. 
\end{eqnarray}
Therefore, using the constraint 
and the formula,  
\begin{eqnarray}
{\rm det}(g+\delta g) 
 &=&{\rm det}[g(1+g^{-1}\delta g)] 
\nonumber \\ 
 &=&-\rho^2\left(1+{\rm tr}(g^{-1}\delta g)+{\rm det}(g^{-1}\delta g)\right) 
\nonumber \\
 &\simeq&-\rho^2\left(1+{\rm tr}(g^{-1}\delta g)\right) \,, 
\end{eqnarray}
we can see in the next order that the metric has to satisfy the constraint 
\begin{eqnarray}
\sum_{I=t,\phi,\psi}\frac{\delta g_{II}}{g_{II}}=0 \,,
\end{eqnarray}
which is the same constraint as in the vacuum case~\cite{Harmark}. 
Then, the asymptotic flatness, eq.~(\ref{condi:AF}), requires that 
the limiting behavior of the metric be 
\begin{eqnarray}
&&g_{tt} \simeq -1 + \frac{4M_{ADM}}{3\pi}\frac{1}{\sqrt{\rho^2+z^2}} 
                + {\cal O}\left(\frac{1}{\rho^2+z^2}\right) \,, 
\label{gtt}
\\
&&g_{t\phi} \simeq -\frac{J_\phi}{\pi}\frac{\sqrt{\rho^2+z^2}-z}{\rho^2+z^2} 
                   +{\cal O}\left(\frac{1}{\rho^2+z^2}\right) \,, 
\label{gtphi}
\\ 
&&g_{t\psi}\simeq  - \frac{J_\psi}{\pi}\frac{\sqrt{\rho^2+z^2}+z}{\rho^2+z^2} 
                   +{\cal O}\left(\frac{1}{\rho^2+z^2}\right) \,, 
\label{gtpsi}
\\
&&\lambda_{\phi\phi}
  \simeq (\sqrt{\rho^2+z^2}-z)
         \left(1+\frac{2(M_{ADM}+\eta)}{3\pi\sqrt{\rho^2+z^2}}
                +{\cal O}\left(\frac{1}{\rho^2+z^2}\right)
         \right) \,, 
\label{gphiphi}
\\
&&\lambda_{\psi\psi}\simeq 
                    (\sqrt{\rho^2+z^2}+z)
          \left(1+\frac{2(M_{ADM}-\eta)}{3\pi\sqrt{\rho^2+z^2}}+{\cal O} 
                \left(\frac{1}{\rho^2+z^2}\right)
          \right) \,,
\label{gpsipsi}
\\ 
&&\lambda_{\phi\psi}\simeq 
          \zeta\frac{\rho^2}{(\rho^2+z^2)^{3/2}} 
          +{\cal O}\left(\frac{1}{\rho^2+z^2}\right) \,, 
\\
&&g_{\rho\rho}=g_{zz}\simeq \frac{1}{2\sqrt{\rho^2+z^2}} 
                           +{\cal O}\left(\frac{1}{\rho^2+z^2}\right) \,, 
\label{grhorho}
\end{eqnarray}
where the constant $M_{ADM}$ denotes the conserved mass defined by 
eq.~(\ref{def:Mass}) and $J_\phi$ and $J_\psi$ the angular momenta, 
defined by eq.~(\ref{def:Ja}). 
Here $\eta$ is a constant that comes from 
gauge degrees of freedom in the choice of the coordinate $z$, 
i.e., degrees of freedom with respect to shift translation $z\to z+\alpha$.  
(This gauge freedom exists even after the gauge freedom of 
the conjugate coordinate, $\rho$, is fixed at infinity.) 
Since in our proof we choose the coordinate $z$ such that the horizons 
are located at the interval $[-k^2,k^2]$ for two configurations 
$M_{[0]}$ and $M_{[1]}$, we choose the same values of $\eta$ 
for the two solutions. 

\medskip
The left-hand side of the Einstein-Maxwell equation behaves as 
${\cal O}((\rho^2+z^2)^{-1})$ in a neighborhood of the infinity. 
The energy-momentum tensor of the Maxwell field must also behave as 
${\cal O}((\rho^2+z^2)^{-1})$. Hence from the asymptotic flatness, 
the gauge potential must behave as 
\begin{eqnarray}
&&A_t\simeq \frac{2Q}{\pi\sqrt{\rho^2+z^2}} 
       + {\cal O}\left(\frac{1}{\rho^2+z^2}\right) \,, 
\label{eq:At} \\
&&\psi_\phi\simeq {\cal O}\left(\frac{1}{\sqrt{\rho^2+z^2}}\right) \,, 
\label{eq:Aphi} \\
&&\psi_\psi \simeq {\cal O}\left(\frac{1}{\sqrt{\rho^2+z^2}}\right) \,. 
\label{eq:Apsi} 
\end{eqnarray} 

\medskip
Next, we derive the behavior of $\mu$ and $\omega_a$ near infinity. 
The magnetic potential, $\mu$, is determined by eq.~(\ref{eq:mu}). 
From eqs.~(\ref{eq:Aphi}) and (\ref{eq:Apsi}), the second term in the 
right-hand side of eq.~(\ref{eq:mu}) behaves as 
${\cal O}((\rho^2+z^2)^{-1})$. The leading term $\mu^{(0)}$, 
where $\mu\simeq\mu^{(0)}+{\cal O}((\rho^2+z^2)^{-1/2})$, is derived 
from the equations 
\begin{eqnarray}
\mu_{,z}^{(0)}\simeq -\frac{\rho}{\sqrt{3}}A_{t,\rho} \,, \quad 
\mu_{,\rho}^{(0)}\simeq \frac{\rho}{\sqrt{3}}A_{t,z} \,. 
\end{eqnarray}
Using the asymptotic behavior (\ref{eq:At}) of the gauge field $A_t$, 
we obtain 
\begin{eqnarray}
\mu^{(0)}= \frac{2Qz}{\pi\sqrt{3}\sqrt{\rho^2+z^2}} \,.  
\end{eqnarray}
The twist potential, $\omega_a$, is determined by 
eq.~(\ref{eq:twistpotential}). The second term behaves as 
${\cal O}((\rho^2+z^2)^{-1})$. Hence, the leading term $\omega_{a}^{(0)}$, 
where $\omega_a\simeq\omega_a^{(0)}+{\cal O}((\rho^2+z^2)^{-1/2})$, 
is derived from the equations 
\begin{eqnarray}
\omega_{a,z}^{(0)}&\simeq&\frac{\tau}{\rho}\lambda_{ab}a^b{}_{t,\rho} \,,
\label{eq:aa1}\\
\omega_{a,\rho}^{(0)}&\simeq&-\frac{\tau}{\rho}\lambda_{ab}a^b{}_{t,z} \,.
\label{eq:aa2}
\end{eqnarray}
The functions $a^a{}_t$ behaves as 
\begin{eqnarray}
&&a^\phi{}_t=\frac{\lambda_{\phi\psi}g_{t\psi}
            -\lambda_{\psi\psi}g_{t\phi}}{\tau}\simeq 
            -\frac{J_\phi}{\pi}\frac{1}{\rho^2+z^2} \,, 
\\
&&a^\psi{}_t=\frac{\lambda_{\phi\psi}g_{t\phi}
              -\lambda_{\phi\phi}g_{t\psi}}{\tau}
            \simeq  
              -\frac{J_\psi}{\pi}\frac{1}{\rho^2+z^2} \,. 
\end{eqnarray}
Therefore, solving eqs.~(\ref{eq:aa1}) and (\ref{eq:aa2}), we obtain
\begin{eqnarray}
\omega_{\phi}^{(0)}&=& \frac{J_\phi}{\pi}
                       \left(\frac{\rho^2}{\rho^2+z^2}
                             -\frac{2z}{\sqrt{\rho^2+z^2}} 
                       \right) \,, \\
\omega_{\psi}^{(0)}&=& \frac{J_\psi}{\pi}
                       \left(\frac{\rho^2}{\rho^2+z^2} 
                             -\frac{2z}{\sqrt{\rho^2+z^2}} 
                       \right) \,. 
\end{eqnarray}
%
%
Then, for $\sqrt{\rho^2+z^2} \to \infty$, $\rho\ {\rm tr}\Psi$ behaves as  
\begin{eqnarray}
\rho\ {\rm tr}\Psi
&\simeq&{\cal O}\left(\frac{1}{\rho^2+z^2}\right) \,. 
\end{eqnarray}
Therefore,
\begin{eqnarray}
\rho\ \partial_p{\rm tr}\Psi 
 dS^p\simeq {\cal O}\left(\frac{1}{\sqrt{\rho^2+z^2}}\right) \,.
\end{eqnarray}

\medskip 
Thus, we find from (i)--(iv) that the boundary integral, 
eq.~(\ref{eq:integral}), vanishes on each rod and the infinity. 
The deviation matrix, $\Psi$, is constant and has the asymptotic behavior, 
$\Psi\to 0$. Therefore, $\Psi$ vanishes over $\Sigma$, and the two 
configurations, $M_{[0]}$ and $M_{[1]}$, coincide with each other. 
Furthermore, as shown in Appendix~\ref{sec:solution}, 
the boundary conditions derived above are the same as the limiting behavior 
of the Cveti{\v c}-Youm solution with equal charges at each corresponding boundary. 
Therefore, the data $M_{[0]}$ (and now equivalently $M_{[1]}$) must also be 
the same as the corresponding matrix to the five-dimensional Cveti{\v c}-Youm solution with equal charges. This completes our proof for the uniqueness theorem.

\section{Summary}\label{sec:conclusion} 

We have shown the uniqueness theorem which states that 
in five-dimensional Einstein-Maxwell-Chern-Simons theory, an asymptotically 
flat, stationary charged rotating black hole with finite temperature is 
uniquely specified by its asymptotic conserved charges and therefore is 
described by the five-dimensional Cveti{\v c}-Youm solution with equal charges, if (1) it admits 
two independent axial Killing symmetries and (2) the topology of the event 
horizon cross-section is spherical. 
%
%
Our theorem generalizes the uniqueness theorem for spherical black 
holes in five-dimensional vacuum Einstein gravity \cite{MI} to 
the case of EMCS theory. 
In our proof, in addition to the symmetry-assumption~(1), the Chern-Simons 
term in the theory, eq.~(\ref{eq:action}), plays an important role 
to reduce the system into a non-linear sigma model with desired symmetry 
property, $G_{2(2)}/SO(4)$, as discussed in~\cite{MO,MO2}.  
Then, having this symmetry property on the target space, 
we have obtained the matrix representation of \cite{BCCGSW},  
in which our system is completely determined by $G_{2(2)}/SO(4)$ coset 
matrix $M$. We then derived the Mazur identity, and used the identity 
to show that if two solutions, i.e., two matrices, $M_{[0]}$ and $M_{[1]}$, 
satisfy the same boundary conditions (imposed at infinity, on two rotational 
axis, and on the horizon), then the solutions $M_{[0]}$ and $M_{[1]}$ 
must coincide with each other. We have shown that our boundary conditions 
(the asymptotic flatness and the regularity) are the same as the limiting 
behavior of the Cveti{\v c}-Youm solution.

\medskip 
In the present theorem, we restrict attention to topologically spherical 
black holes by the assumption~(2). 
Our theorem can be generalized to 
the case of charged rotating black ring solutions by imposing certain 
additional conditions.  
We first note that under the same symmetry condition~(1), the analysis 
in Sec.~\ref{sec:metric}, \ref{sec:mazur} and \ref{sec:coincidence} apply 
also for black ring solutions (if exist) in EMCS theory. 
(See~\cite{EEF} for such a ring solution.)
%
%
The only difference from the spherical 
black hole case arises in the boundary value analysis. 
Now we also note that asymptotically flat, five-dimensional black ring 
solutions that satisfy the symmetry assumption~(1) have the following rod 
structure: 
(i) $[c,\infty], \ v=(0,1,0)$, 
(ii) $[ck^2,c],\ v=(0,0,1)$,   
(iii) $[-ck^2,ck^2]$, and 
(iv) $[-\infty,-ck^2],\ v=(0,0,1)$, where $c>0, k^2<1$ and $v$'s are 
eigenvectors with respect to a zero eigenvalue of the three-dimensional 
matrix $g_{IJ}$ for each segment. 
It should be noted that we are not concerned with a lens space 
throughout discussion here, and therefore the only non-trivial rod data 
are given by rod intervals. Then, after fixing the scale $c$, 
one can completely specify the rod data in terms of $k^2$.  
The finite spacelike rod (ii) is the main difference from the rod structure 
for topologically spherical black holes considered in 
Sec.~\ref{sec:coincidence}. 
We believe that by appropriately specifying rod structure, one can 
determine the topology of the horizon, as well as the topology of 
black hole exterior region. In this respect, it has recently been 
shown \cite{HY08} that the topology and symmetry structure of the black hole 
spacetime can be completely determined in terms of {\em rod-intervals}, 
which is similar to but somewhat different from the rod-structure of 
Harmark \cite{Harmark}. 
%
%
In the charged black ring case, a dipole charge may also play a role. 
These issues deserve further study.

\bigskip

\section*{Acknowledgments} 
We would like to thank P. Figueras for comments and discussions on 
possible generalization of the present results to the black ring case. 
We also would like to thank S. Hollands for discussions 
concerning the topology of the black hole exterior region. 
ST is supported by the JSPS under Contract No. 20-10616.

\appendix

\section{Maxwell field with symmetries}\label{sec:A}
Let $F$ denote the stationary and axisymmetric Maxwell field, i.e., 
that satisfies 
\begin{eqnarray}
{\cal L}_{\xi_I} F=0 \,, 
\end{eqnarray}
with $\xi_I\ (I=\phi,\psi,t)$ being commuting Killing vector fields 
for the axial-symmetries and the stationary symmetry,  
discussed in sec.~\ref{subsec:II-B}.  
From the Maxwell equation, $dF=0$, and the identity
\begin{eqnarray}
di_{\xi_I}i_{\xi_J}&=&-i_{\xi_I}{\cal L}_{\xi_J}
                      +i_{\xi_J}{\cal L}_{\xi_I}+i_{\xi_I}i_{\xi_J}d \,, 
\end{eqnarray} 
we have 
\begin{eqnarray}
di_{\xi_I}i_{\xi_J}F &=& -i_{\xi_I}{\cal L}_{\xi_J}F 
                         +i_{\xi_J}{\cal L}_{\xi_I}F 
                         +i_{\xi_I}i_{\xi_J}dF 
\nonumber \\ 
                     &=&0 \,. 
\end{eqnarray}
Similarly, using the identity 
\begin{eqnarray}
di_{\xi_I}i_{\xi_J}i_{\xi_K}=i_{\xi_I}i_{\xi_J}{\cal L}_{\xi_K}-i_{\xi_I}i_{\xi_K}{\cal L}_{\xi_J}+i_{\xi_J}i_{\xi_K}{\cal L}_{\xi_I}-i_{\xi_I}i_{\xi_J}i_{\xi_K}d \,,
\end{eqnarray}
we have 
\begin{eqnarray}
di_{\xi_I}i_{\xi_J}i_{\xi_K}*F
 &=&i_{\xi_I}i_{\xi_J}{\cal L}_{\xi_K}*F
   -i_{\xi_I}i_{\xi_K}{\cal L}_{\xi_J}*F 
   +i_{\xi_J}i_{\xi_K}{\cal L}_{\xi_I}*F
   -i_{\xi_I}i_{\xi_J}i_{\xi_K}d*F 
\nonumber \\
 &=&i_{\xi_I}i_{\xi_J}*{\cal L}_{\xi_K}F 
   -i_{\xi_I}i_{\xi_K}*{\cal L}_{\xi_J}F 
   +i_{\xi_J}i_{\xi_K}*{\cal L}_{\xi_I}F 
   -i_{\xi_I}i_{\xi_J}i_{\xi_K}d*F 
\nonumber \\ 
 &=&\frac{1}{\sqrt{3}}i_{\xi_I}i_{\xi_J}i_{\xi_K} F\wedge F 
\nonumber \\
 &=& 0 \,. 
\end{eqnarray}
Therefore, $F(\xi_I,\xi_J)$ and $(*F)(\xi_I,\xi_J,\xi_K)$, are constant. 
Since they vanish, at least, on rotation axes, these imply 
\begin{eqnarray}
&&F(\xi_I,\xi_J)=0 \,,\\
&&(*F)(\xi_I,\xi_J,\xi_K)=0 \,.
\end{eqnarray}
In terms of the coordinates $(t,\phi,\psi,\rho,z)$, these can be written as 
\begin{eqnarray}
&&F_{t\phi}=F_{t\psi}=F_{\phi\psi}=0 \,, \\ 
&&F_{\rho z}=0 \,. \label{eq:Frz}
\end{eqnarray}
Then, from (\ref{eq:Frz}), using the gauge degrees of freedom, 
$A \to A-d\chi$, with the function $\chi$ satisfying 
$A_\rho=\chi_{,\rho} \,, \quad A_\theta=\chi_{,\theta}$ we can show 
\begin{eqnarray}
A_\rho=A_z=0 \,. 
\end{eqnarray} 

\section{Cveti{\v c}-Youm solution with equal charges} 
\label{sec:solution} 
Here we present the asymptotically flat stationary charged rotating 
black hole solution in five-dimensional Einstein-Maxwell-Chern-Simons 
theory, found by Cveti\v{c} et al~\cite{CY96}. The solution has three 
mutually commuting Killing vectors that generate 
isometries ${\Bbb R} \times U(1) \times U(1)$, and spherical 
topology of the horizon cross-sections. 
We observe that the limiting behavior of relevant scalar functions 
of the solution, which correspond to $\Phi^A$, are in perfect accordance 
with our general boundary conditions discussed in Sec.~\ref{sec:coincidence}. 

\medskip 
The metric and the gauge potential in \cite{CY96,CLP,CCLP} are given, 
respectively, by 
\begin{eqnarray}
ds^2&=& 
     -dt^2-\frac{2q}{\tilde \rho^2}\nu(dt-\omega)
     +\frac{f}{\tilde\rho^4}(dt-\omega)^2 
     +\frac{\tilde \rho^2r^2}{\Delta}dr^2+\tilde\rho^2d\theta^2 
\nonumber \\
   &+&(r^2+a^2)\sin^2\theta d\phi^2+(r^2+b^2)\cos^2\theta d\psi^2 \,, 
\label{eq:solution}
\end{eqnarray}
and
\begin{eqnarray}
A=\frac{\sqrt{3}q}{\tilde\rho^2}(dt-\omega) \,,\label{eq:gauge}
\end{eqnarray}
where
\begin{eqnarray}
&&\nu=b\sin^2\theta d\phi+a\cos^2\theta d\psi \,,\\
&&\omega=a\sin^2d\phi+b\cos^2\theta d\psi \,, \\
&&f=2m\tilde\rho^2-q^2 \,, \\
&&\Delta=(r^2+a^2)(r^2+b^2)+q^2+2ab q-2mr^2 \,, \\
&&\tilde\rho^2=r^2+a^2\cos^2\theta+b^2\sin^2\theta \, .  
\end{eqnarray}
The scalar fields $\Phi^A=(\lambda_{ab},\omega_a,\psi_a,\mu)$ 
for the solution (\ref{eq:solution}) and (\ref{eq:gauge}) are computed as 
\begin{eqnarray}
&&\lambda_{\phi\phi}=\frac{2q}{\tilde\rho^2}ab \sin^4\theta
       +\frac{f}{\tilde\rho^4}a^2\sin^4\theta+(r^2+a^2)\sin^2\theta \,, 
\\
&&\lambda_{\psi\psi}=\frac{2q}{\tilde\rho^2}ab \cos^4\theta 
       +\frac{f}{\tilde\rho^4}b^2\cos^4\theta+(r^2+b^2)\cos^2\theta \,, 
\\
&&\lambda_{\phi\psi}=\frac{q}{\tilde\rho^2}(a^2+b^2)
  \cos^2\theta\sin^2\theta+\frac{f}{\tilde\rho^4}ab\cos^2\theta\sin^2\theta \,,
\end{eqnarray} 
\begin{eqnarray} 
\omega_\phi&=&\frac{(2am+bq)(-4\cos2\theta+\cos4\theta)}{8} 
\nonumber\\
         &&-\frac{2(a^2-b^2)(2aq^2+(2am+bq)F)\cos^2\theta\sin^4\theta}{F^2} \,,
\\
\omega_\psi&=&-\frac{(2bm+aq)(4\cos2\theta+\cos4\theta)}{8} 
\nonumber\\
         &&-\frac{2(a^2-b^2)(2bq^2+(2bm+aq)F)\cos^4\theta\sin^2\theta}{F^2} \,,
\end{eqnarray} 
\begin{eqnarray}
\psi_\phi&=&-\frac{qa\sin^2\theta}{\tilde\rho^2} \,,\\
\psi_\psi&=&-\frac{qb\cos^2\theta}{\tilde\rho^2} \,, \\
\mu&=&\frac{1}{2}q\cos2\theta-\frac{2(b^2-a^2)q\cos^2\theta\sin^2\theta}{F} \,,
\end{eqnarray}
where the function $F$ is defined by 
\begin{eqnarray}
F=a^2+b^2+2r^2+(a^2-b^2)\cos2\theta \,. 
\end{eqnarray}

\medskip 
Let us introduce the coordinates $(\rho,z)$ defined by 
\begin{eqnarray}
\rho=\frac{1}{2}\sqrt{\Delta}\sin2\theta \,,\quad 
 z =\frac{2r^2+a^2+b^2-2m}{4}\cos2\theta \,. 
\end{eqnarray}
Then, the base space $\Sigma=\{(\rho,z)|\ \rho>0,\ -\infty<z<\infty \}$ 
has four boundaries, which exactly correspond to the four boundaries 
discussed in Sec.~\ref{sec:coincidence}: Namely, 
(i) $\phi$-invariant plane, i.e., the plane which is invariant under 
the rotation with respect to the Killing vector field 
$\partial/\partial \phi$: 
$\partial \Sigma_\phi=\{(\rho,z)|\rho=0,k^2<z<\infty \}$, 
(ii) Horizon: 
$\partial \Sigma_{\cal H}=\{(\rho,z)|\ \rho=0,-k^2<z<k^2\}$, 
(iii) $\psi$-invariant plane, i.e., the plane which is invariant 
under the rotation with respect to the Killing vector field 
$\partial /\partial \psi$: 
$\partial \Sigma_\psi=\{(\rho,z)|\rho=0,-\infty<z<-k^2\}$, and 
(iv) Infinity: 
$\partial \Sigma_\infty=\{(\rho,z)|\sqrt{\rho^2+z^2}\to\infty 
  \ {\rm with}\ z/\sqrt{\rho^2+z^2}\ {\rm finite}  \}$, 
where the constant $k^2$ is given by 
\begin{eqnarray}
 k^2=\frac{\sqrt{(2m-a^2-b^2)^2-4(ab+q)^2}}{4} \,. 
\end{eqnarray}

\medskip 
Let us examine the behavior of the scalar fields on each boundary. 

\medskip 
\noindent 
(i) Near the {$\phi$-invariant plane} $\partial \Sigma_\phi$, 
each scalar field behaves as 
\begin{eqnarray}
&&\lambda_{\phi\phi}\simeq {\cal O}(\rho^2)\,, \quad 
\lambda_{\psi\psi}\simeq {\cal O}(1) \,, \quad 
\lambda_{\phi\psi}\simeq {\cal O}(\rho^2) \,, 
\\
&& \omega_\phi \simeq -\frac{3}{8}(2am+bq) 
                 + {\cal O}(\rho^2),\, \quad 
   \omega_\psi \simeq -\frac{5}{8}(2bm+aq)+{\cal O}(\rho^2) \,, 
\\
&& \psi_\phi \simeq {\cal O}(\rho^2) \,, \quad 
   \psi_\psi \simeq {\cal O}(1) \,, \quad 
   \mu \simeq \frac{1}{2}q+{\cal O}(\rho^2) \,.
\end{eqnarray}

\smallskip 
\noindent 
(ii) Near the horizon $\partial \Sigma_{\cal H}$, the scalar fields 
behave as 
\begin{eqnarray}
&&\lambda_{ab} \simeq {\cal O}(1) \,, \quad 
  \omega_{a} \simeq {\cal O}(1) \,, \\
&&\psi_{a} \simeq {\cal O}(1) \,, \quad  
  \mu \simeq {\cal O}(1) \,.
\end{eqnarray}

\smallskip 
\noindent 
(iii) Near the {$\psi$-invariant plane} $\partial \Sigma_\psi$, 
each potential behaves as 
\begin{eqnarray}
&&\lambda_{\phi\phi}\simeq {\cal O}(1) \,, \quad 
  \lambda_{\psi\psi}\simeq {\cal O}(\rho^2) \,, \quad 
  \lambda_{\phi\psi}\simeq {\cal O}(\rho^2) \,, 
\\
&&\omega_\phi \simeq \frac{5}{8}(2am+bq)+{\cal O}(\rho^2)\,, \quad 
  \omega_\psi \simeq\frac{3}{8}(2bm+aq)+{\cal O}(\rho^2)\,, 
\\
&&\psi_\phi \simeq {\cal O}(1) \,, \quad 
  \psi_\psi \simeq {\cal O}(\rho^2) \,, \quad 
  \mu \simeq -\frac{q}{2}+{\cal O}(\rho^2) \,. 
\end{eqnarray}

\smallskip
\noindent 
(iv) In the neighborhood of infinity $\partial\Sigma_\infty$, 
the behavior of the potentials becomes 
\begin{eqnarray}
&&\lambda_{\phi\phi}\simeq 
           (\sqrt{\rho^2+z^2}-z)
           \left(1+\frac{a^2}{2\sqrt{\rho^2+z^2}}\right)
           +{\cal O}\left(\frac{1}{\rho^2+z^2}\right) \,, \\
&&\lambda_{\psi\psi}\simeq 
           (\sqrt{\rho^2+z^2}+z)
           \left(1+\frac{2m-a^2}{2\sqrt{\rho^2+z^2}}\right)
           +{\cal O}\left(\frac{1}{\rho^2+z^2}\right) \,, \\ 
&&\lambda_{\psi\phi}\simeq 
           \frac{(a^2q+b^2q+2abm)\rho^2}{8(\rho^2+z^2)^{3/2}}
           +{\cal O}\left(\frac{1}{\rho^2+z^2}\right) \,, \\ 
&& \omega_{\phi}\simeq 
           \frac{1}{8}(2am+bq)(-4\cos2\theta+\cos4\theta) 
           +{\cal O}\left(\frac{1}{\sqrt{\rho^2+z^2}}\right) \,, \\
&&\omega_{\psi}\simeq-\frac{1}{8}(2bm+aq)(4\cos2\theta+\cos4\theta) 
             +{\cal O}\left(\frac{1}{\sqrt{\rho^2+z^2}}\right) \,, \\
&&\psi_\phi\simeq 
           -\frac{qa(\sqrt{\rho^2+z^2}-z)}{4(\rho^2+z^2)} 
           + {\cal O}\left(\frac{1}{\rho^2+z^2}\right) \,,\\
&&\psi_\psi\simeq 
            -\frac{qb(\sqrt{\rho^2+z^2}+z)}{4(\rho^2+z^2)} 
            +{\cal O}\left(\frac{1}{\rho^2+z^2}\right) \,, \\ 
&&\mu \simeq 
             \frac{qz}{2\sqrt{\rho^2+z^2}} 
            +{\cal O}\left(\frac{1}{\sqrt{\rho^2+z^2}}\right) \,. 
\end{eqnarray}

\end{document}